\documentstyle[12pt]{article}        
\newlength{\dinwidth}
\newlength{\dinmargin}
\def\lsim{\mathrel{\rlap{\lower4pt\hbox{\hskip1pt$\sim$}}
    \raise1pt\hbox{$<$}}}                
\def\gsim{\mathrel{\rlap{\lower4pt\hbox{\hskip1pt$\sim$}}
    \raise1pt\hbox{$>$}}}                
\def\ra{\rightarrow}
\def\tb{\tan \beta}
\def\bi{\bibitem}

\def\eg{ {\it e.g.} }
\def\ie{ {\it i.e.} }
\begin{document}
\begin{flushright}
IFT 21-96\\ [1.5ex]
{\large \bf hep-ph/9609477 } \\ [1.5ex]
\end{flushright}
\vspace*{1cm}
\begin{center}  \begin{Large} \begin{bf}
Higgs search at HERA\footnote{Contribution to ``Future Physics at
HERA'', 1995--1996, Hamburg, DESY}\\
  \end{bf}  \end{Large}
  \vspace*{5mm}
  \begin{large}
Maria Krawczyk\\
  \end{large}
\end{center}
\begin{center}
Institute ~of ~Theoretical ~Physics,\\
~Warsaw ~University, ~Ho\.za 69,
00 681 ~Warsaw, ~Poland\\
\end{center}
\begin{quotation}
\noindent
{\bf Abstract:}
Present data do not rule out a light neutral Higgs particle
with mass below 40--50 GeV
in the framework of 2HDM with
$ \tb \sim $ 20-30.
The  promising possibility of searching for
a light  Higgs particle in such a scenario in photoproduction at HERA
collider is discussed.
  For the MSSM there is only a very small chance to observe the
Higgs sector, and only for limited mass range $\sim$ 45-50 GeV
and with large $\tb$.

\end{quotation}
\section{Introduction}
The possibilities of Higgs searches at the HERA collider have been
studied in the first HERA Workshop in 1987 \cite{hera88}. It was found that it
`is (almost) impossible'  to observe the Standard Model (SM) Higgs.
A similar conclusion for the SM scalar Higgs search at HERA,
even with an upgraded luminosity and/or proton energy,
can be found in the contribution to this Workshop \cite{hera96}.

Non-minimal Higgs boson production at HERA has also been
investigated during the HERA Workshop'87 as well as
in other papers \cite{grz,bk1,bk}. It was found that
photon-gluon fusion into  $b {\bar b} h (A)$ may be an important
production mechanism of Higgs bosons in the two Higgs doublet
extension of the SM at HERA.  Also other
subprocesses with Higgs boson bremsstrahlung,
namely those with
the resolved photon in the initial state,
are important at HERA \cite{bk1}. Another production mechanism
is gluon-gluon fusion via a quark loop, where the Higgs particle
is produced in resonance. For large $\tb$ and Higgs masses below 30 GeV
this process  dominates the production cross section
over $\gamma g$ fusion \cite{bk}.

According to LEP I data, Higgs bosons in the MSSM have to be heavier than
45 GeV, therefore their production rate at the HERA collider is rather
small.  For the mass of $h$ and $A$ equal to 45 GeV
both the $\gamma g$ and the $gg$ cross sections
for
$ \tb $=30 are $\sim$ 5 fb. When adding the similar contribution from
WW fusion into $h$ \cite{hera96},
one may expect 20-30 events to be produced at HERA
with an integrated luminosity of 1 fb$^{-1}$.

The situation is quite different in the non-supersymmetric version
of the two Higgs doublet extension of the SM (the so called general
two Higgs doublet model - 2HDM)
since in this model  one light neutral Higgs boson with mass
below 40-50 GeV  may still  exist,
moreover with large coupling to
$ \tau $  and  $b
$ -quark.
In this case there is a good
chance to study the Higgs sector at HERA,
with one thousand events expected for a Higgs mass of 5 GeV and
$ \tb$=30, assuming
${\cal L}_{ep}$=250 pb$^{-1}$.

Below we present the status of this model, \ie, the
2HDM with a light Higgs boson, and in the next section we
discuss  the possibility to perform a Higgs boson search
 at HERA, focusing mainly on the gluon-gluon fusion production
via a quark loop. Existing limits from LEP I for the coupling of
a light neutral Higgs boson in the 2HDM are rather weak.  Also present
data on $(g-2)$ for the muon improve only slightly the limits on  $\tb$
for a Higgs mass below 2 GeV.
Therefore it is extremely important to check if  more stringent
limits can be obtained from  HERA measurements.
The combined exclusion plot showing the potential
of HERA measurement is presented in Sec.4.  Sec.5
contains our conclusion.

\section{Status of the 2HDM with a Light Neutral Higgs}
The mechanism of spontaneous symmetry breaking  proposed as
the source of mass for the gauge and fermion fields in the
SM leads to a neutral scalar particle, the minimal Higgs boson.
According to  the LEP I data,
based on the Bjorken process $e^+e^- \ra H Z^*$,
it should be heavier than 66 GeV \cite{hi}.
A possible extension of the SM is to include
a second Higgs doublet to the symmetry breaking
mechanism. In the two Higgs doublet models
the observed Higgs sector is enlarged to five scalars: two
neutral Higgs scalars (with masses $M_H$ and $M_h$ for the heavier and
lighter particle, respectively), one neutral pseudoscalar
($M_A$), and a pair of charged Higgses ($M_{H^\pm}$). The
neutral Higgs scalar couplings to quarks, charged leptons and gauge
bosons are
modified with respect to analogous couplings in SM by factors that
depend on additional parameters: $\tan\beta$, which is the ratio
of the vacuum expectation values of the Higgs doublets
$v_2/v_1$, and the mixing angle in the neutral Higgs sector $\alpha$.
Also further new couplings appear, e.g. $Zh (H) A$ and $ZH^+ H^-$.

In the following we will focus on the appealing version of the models
with two doublets ("Model II") where one Higgs doublet
with vacuum expectation value $v_2$ couples only to the "up"
components of fermion doublets while the other one couples only
to the "down" components \cite{hunter}.
({In particular,  fermions couple to the pseudoscalar $A$
with a strength  proportional to $(\tan \beta)^{\pm1}$
whereas the coupling of the fermions to the scalar $h$
goes as $\pm(\sin \alpha/\cos \beta)^{\pm1}$, where the sign
$\pm$  corresponds to  the isospin $\mp$1/2 components}).
In such a model FCNC processes are absent and the $\rho$
parameter retains its SM value at the tree level. Note that in
such a scenario the large ratio $v_2/v_1 \sim m_{top}/m_b\gg 1$
is naturally expected.

The well known supersymmetric model (MSSM) belongs to this  class.
In the MSSM there are additional relations among the parameters required by
supersymmetry, leaving only two free parameters (at the tree level)
e.g. $M_A$ and $\tb$. In the general case, denoted 2HDM, masses and
parameters $\alpha$ and $\beta$ are not constrained.
Therefore the same experimental data may lead to very distinct
consequences  depending on which version of the two Higgs doublet
extension of the SM, supersymmetric or nonsupersymmetric, is considered.

\subsection  {Present constraints on the 2HDM from LEP I.}
Important constraints on the parameters of the two Higgs doublet
extensions of the SM were obtained in the precision measurements at LEP I.
The current mass limit on the {\it {charged}} Higgs boson
$M_{H^{\pm}}$=44 GeV was obtained at LEP I \cite{sob} from the
process $Z \ra H^+H^-$, which is independent of the parameters
$\alpha$ and $\beta$. (Note that in the MSSM version one expects
$M_{H^{\pm}} > M_W$). For the {\it {neutral}} Higgs particles $h$ and
$A$ there are two main and complementary sources  of information at
LEP I. One is the Bjorken process $Z \ra Z^*h$ which constrains
$g_{hZZ}^2 \sim \sin^2(\alpha-\beta)$, for $M_h$ below 50-60 GeV.
The second process is $Z\ra hA$, constraining $g_{ZhA}^2 \sim
\cos^2(\alpha-\beta)$ for $M_h+M_A\lsim M_Z$. Results on
$\sin^2(\alpha-\beta)$ and $\cos^2(\alpha-\beta)$ can be translated
into limits on the neutral Higgs bosons masses $M_h$ and $M_A$.
In the MSSM, due to th additional relations among the parameters,
the above data allow to draw limits for the masses
of {\it {individual}} particles: $M_h\ge 45$ GeV for any $\tan \beta $
 and $M_A \ge$ 45 GeV for $\tan\beta \ge$1 \cite{susy,hi}.
In the general 2HDM the implications are quite different, here
only the large portion of the ($M_h$,$M_A$) plane,
where {\it {both}} masses are in the range between
0 and $\sim$50 GeV, is excluded \cite{lep}.

The third basic process to search for a neutral Higgs particle at
LEP I is the Yukawa process, $\ie$ the bremsstrahlung production
of a neutral Higgs boson $h(A)$ from a heavy fermion: $e^+e^-
\rightarrow f {\bar f} h(A)$, where $f$ means here {\it b} quark
or $\tau$ lepton \cite{gle,pok,kk}. {{ A new analysis of the Yukawa
process by the ALEPH collaboration \cite{alef} led to
an exclusion plot on $\tb$ versus the pseudoscalar mass, $M_A$. (The
analysis by the L3 collaboration is also in progress { \cite{l3prep}}.).
However, the obtained limits are rather weak allowing for the existence
of a light $A$ with mass below 10 GeV with $\tb$ = 20--30, for $M_A$=40
GeV $\tb$ till 100 is allowed! For the mass range above 10 GeV,
similar exclusion limits should in principle hold also for a
scalar $h$ when replacing the coupling $\tb \ra \sin\alpha/
\cos\beta$. However, one would expect larger differences in the lower
mass region,  where the production rate at the same value of coupling
for the scalar is considerably larger than for the pseudoscalar. More
stringent limits should be obtained there\cite{kk}.

In the following we will study the 2HDM assuming
that one  light Higgs particle  may exist.
Moreover we will  assume  according to  LEP I data
the following mass relation between the lightest
neutral Higgs particles: $M_h+M_A \gsim M_Z$. We specify the model
further  by choosing particular values for the parameters $\alpha$
and $\beta$ within the present limits from LEP I. Since $\sin^2(
\alpha-\beta)$ was found \cite{lep,hi} to be smaller than 0.1 for
$10\lsim \;  M_h\lsim \;$ 50 GeV, and  even below 0.01 for a lighter
scalar, we simply  take  $\alpha=\beta$. This leads to equal strengths
of the coupling of  fermions to scalars   and pseudoscalars.
Note that then the EW gauge boson couplings to the Higgs scalar $h$ disappear
\footnote{ $A$ does not couple to W and Z \cite{hunter}.}.
For the scenario with large $\tan\beta \sim {\cal O}(m_t/m_b)$ a large
enhancement in the coupling of both $h$ and $A$ bosons to the down-type
quarks and leptons is expected.

Below we present  how one obtains limits on the parameters of the
2HDM from current muon $(g-2)$ data \cite{pres}, also the potential of
the forthcoming E821 experiment \cite{fut} is discussed. (See
Ref.\cite{g22} for details.)

\subsection{Constraints on the 2HDM  from $(g-2)_\mu$}
The present experimental limit on $(g-2)$ for the muon,
averaged over the sign of the muon electric charge, is given by \cite{data}:
$$a_{\mu}^{exp}\equiv{{(g-2)_{\mu}}\over{2}}=1~165 ~923~(8.4)\cdot 10^{-9}.$$
The quantity within parenthesis, $\sigma_{exp}$, refers to the uncertainty
in the last digit.

The theoretical prediction of the SM for this quantity
consists of the QED, hadronic and EW contributions:
$$a_{\mu}^{SM}=a_{\mu}^{QED}+a_{\mu}^{had}+a_{\mu}^{EW}.$$
The recent  SM calculations of $a_{\mu}$
are based on the  QED results from \cite{qed}, the hadronic contributions
obtained in
\cite{mar,mk,jeg,wort,ll} and \cite{hayakawa} and
the EW results from \cite{czar,kuhto}.
The uncertainties in these
contributions differ among themselves considerably
(see below  and in  Ref.\cite{nath,czar,jeg,g22}).
The main discrepancy
is observed  for  the hadronic contribution,
therefore  we will here consider case A,
based on Refs.\cite{qed,ki,mar,mk,ll,czar},
with relatively small error in the hadronic
part. It corresponds to : $a_{\mu}^{SM}$=1~165~918.27 ~(0.76)
$\cdot 10^{-9}$.
 (The results for  case B (Refs.
\cite{ki,jeg,hayakawa,czar}) with the 2 times
 larger error in the hadronic part is discussed in \cite{czar,g22}.)

The room for new physics, like the 2HDM with a light scalar or
a light pseudoscalar, is basically given by the difference between the
experimental data and the theoretical SM prediction:{\footnote{However in
the calculation of $a_{\mu}^{EW}$ the (SM) Higgs scalar contribution is
included (see discussion in \cite{g22}).}} $a_{\mu}^{exp}-a_{\mu}^{SM}\equiv
\delta a_{\mu}$. Below, the difference $\delta a_{\mu}$ for the considered
case, A, is presented together with the error $\sigma$, obtained by adding
the experimental and theoretical errors in quadrature (in $10^{-9}$):
$$
\delta a_{\mu}(\sigma) =   4.73 (8.43)  {\hspace{0.5cm}} {\rm and}
{\hspace{0.5cm}} {\rm lim_{\pm}(95\%)}:
-13.46\le\delta a_{\mu} \le 19.94
$$
One can see that at the 1 $\sigma$ level the difference $\delta a_{\mu}$
can be positive or negative. For the beyond the SM scenario where the
contribution of only {\it {one}} sign is physically accessible
($\ie$ positive or negative $\delta a_{\mu}$), the other sign being
unphysical, the 95\%C.L. limits should be calculated \cite{data}
separately for the positive and for the negative contributions
($lim_{\pm}$(95\%) above ).

The future accuracy of the $(g-2)_{\mu}$ experiments is expected to be
 $\sigma^{new}_{exp}\sim0.4 \cdot 10^{-9}$ or better \cite{fut,czar}.
One also expects an improvement in the calculation of the hadronic contribution
{\footnote {An improvement in the ongoing experiments at low energy
is expected as well.}} such that the total uncertainty will be basically
due to the experimental error. Below we will assume that
the accessible range for the beyond SM contribution
will be smaller by a factor of 20 compared to the present
($lim_{\pm}$95\%) bounds. So, we consider the following option for future
measurements (in $10^{-9}$):
$$
{\rm lim_{\pm}}^{new}(95\%): -0.69\le\delta a_{\mu}^{new} \le 1.00.
$$
The difference $\delta a_{\mu}^{new}$ we now ascribe to the 2HDM contribution,
so we take $\delta a_{\mu}= a_{\mu}^{(2HDM)}$ and $\delta a_{\mu}^{new}
= a_{\mu}^{(2HDM)}$ for present and future $(g-2)_\mu$ data, respectively.
We will consider  two scenarios:
\begin{itemize}
\item {\sl a)} pseudoscalar $A$ is light
\item {\sl b)} scalar $h$ is light.
\end{itemize}

Here we calculate the 2HDM contribution assuming for case {\sl a})
 $  a_{\mu}^{(2HDM)}(M_A)= a_{\mu}^A(M_A)$, whereas for {\sl b}) :
$   a_{\mu}^{(2HDM)}(M_h)= a_{\mu}^h(M_h)$.
This simple approach is based on the LEP I mass limits for charged and
neutral Higgs particles and differs from the full 2HDM predictions,
 studied in Ref.\cite{g22}, significantly for a Higgs mass above about 30 GeV.
Note that the contribution for the scenario {\sl b)} is positive,
whereas for the scenario {\sl a)} it is negative.

The exclusion plots for $\tb$ obtained from present $(g-2)_{\mu}$ data at
95\%C.L. for a light $h$ or  $A$, go beyond those from LEP I for Higgs
masses below 2 GeV. $\tb \sim$ 15 is still allowed for the mass of the
Higgs particle as low as 1 GeV, above 2 GeV $\tb$ is limited to 20.
These results together with others will be presented later in Sec. 4.

It is worth pointing out the unique role of the forthcoming $(g-2)_{\mu}$
measurement in clarifying which scenario of the 2HDM is allowed:
the model with a light scalar or with a light pseudoscalar. If the
$\delta a_{\mu}^{new}$ is positive (negative) then the light pseudoscalar
(scalar) is excluded. Further constraints on the coupling of the allowed
light Higgs particle can be obtained from
HERA, which is very well suited for this task.

\section{Search for a Higgs particle at HERA}
We now study the possibility of light neutral Higgs scalar and/or
pseudoscalar production in a 2HDM at HERA \cite{bk1,bk,ames}.
We limit ourselves to the mass range above 5 GeV, in order to apply
the LO approach. Next order results are in preparation \cite {deb9},
but we expect that even a K-factor
$\sim$ 1.5-2 will not change the results drastically.
The results obtained for 2HDM
hold also for MSSM, provided the proper range of mass
is considered, \ie, above 45 GeV.
The results relevant for SM
can also be obtained from the
2HDM predictions for $h$ production
with $\tb$=1.

 Photon-gluon and gluon-gluon fusions in
photoproduction at HERA are expected to be 
 basic sources of a light Higgs
bosons in 2HDM \cite{bk}.
Note that the $ZZ$ and $WW$ fusions are not relevant
since the pseudoscalar, $A$, does not couple to EW gauge bosons,
and the scalar couplings, which are
proportional to $\sin(\alpha-\beta)$, are put to zero by the
assumed by us condition: $\alpha=\beta$.
(Even if the upper experimental limits for $\sin(\alpha-\beta)$
are used (Sec. 2.1), light Higgs boson production in
$ZZ/WW$ fusion is still effectively suppressed.)

The total cross section for on-shell neutral Higgs boson production
is calculated, along with the rates
for the particular final states $\ra \tau^+ \tau^{-}$ or $bb$.
These decay channels are the most important
since in 2HDM
with  large $\tb$ $h$ and $A$ decay mainly to the heaviest available fermionic
``down-type'' states.
(Details can be found elsewhere, \eg, in \cite{pok,deb12}).

\subsection{Bremsstrahlung of the Higgs boson: $\gamma g\ra b {\bar b}h(A)$
and other related processes.}
At HERA, the photoproduction of neutral Higgs boson
in a 2HDM
via
\begin{equation}
\gamma g \ra b {\bar b} h (A)
\end{equation}
may be substantial \cite{grz,bk}. The total
$ h$ cross section
(integrated over the
$b {\bar b}$ final state)
is presented in Fig.1a.
 Note that this process
also includes
$\gamma b\ra bh(A)$,
as well
the lowest order contributions due to the resolved photon,
{\em i.e.,}
$b {\bar b}\ra h(A)$, $bg \ra h(A)b$,
etc.  These subprocesses   were studied in
Ref. \cite{bk1},  in Fig.2a we present obtained results.
Each of these processes needs an independent analysis of the
background \cite{deb9}. As this work is not yet completed,
we will not use these processes for the
derivation of the exclusion plot in Sec.4.
\begin{figure}[ht]  \label{fig:herab}
\vskip 4.1in\relax\noindent\hskip -0.8in
             \relax{\includegraphics{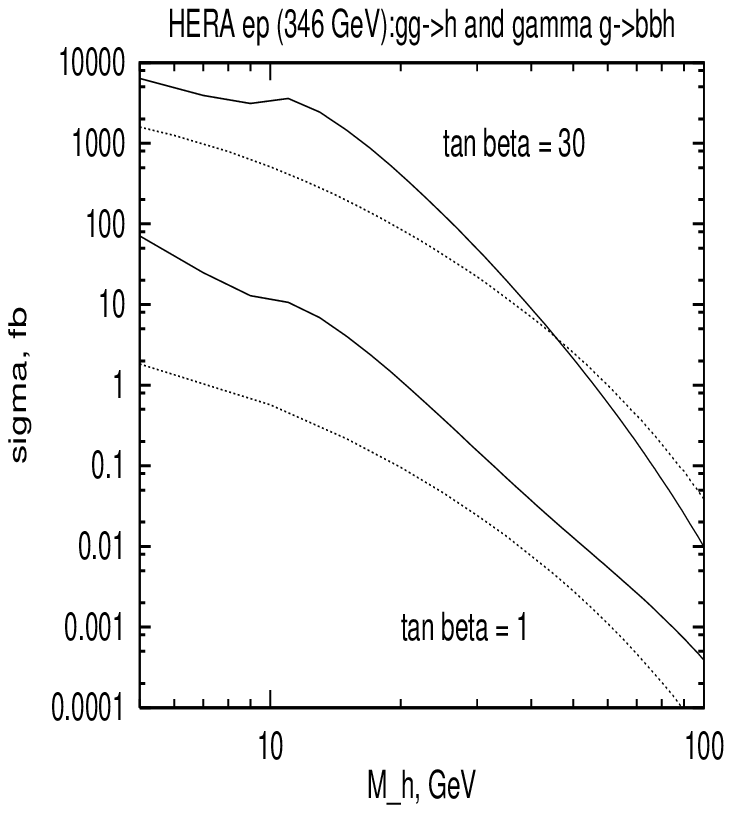}}
             \relax\noindent\hskip 3.5in
              \relax{\includegraphics{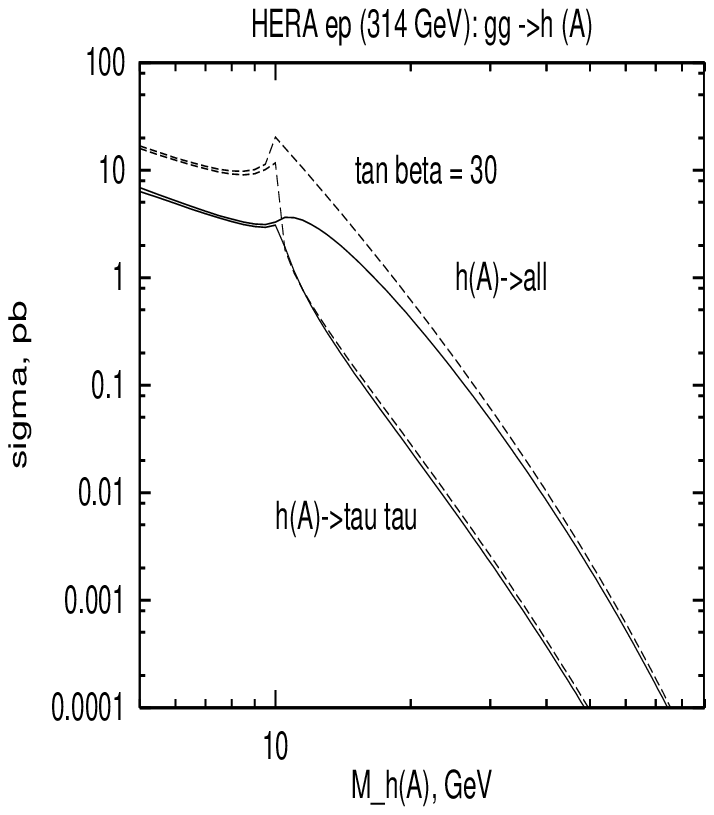}}
\vspace{-12.5ex}
\caption{ { \em
a) The total cross section
for $h$ production at HERA with the upgraded energy.
The $gg$ fusion (solid lines)
and $\gamma g$ fusion  (dotted lines) are considered.
The results for $ \tb $ =30 (upper curves)
and the $ \tb $=1  (lower curves) are shown.
GRV parametrizations were used for both
the photon and the proton [18].
b) The cross sections for $gg$ fusion for the nominal HERA energy (with $\tb$=30).
Total $h$ (solid upper line)
and  $A$ (dashed upper line) production cross-sections are shown, along with
the results for the
$ \tau^+ \tau^- $
final state (solid and dashed lower lines for $h$ and $A$, respectively).}}
\end{figure}
\subsection{Gluon-gluon fusion via a quark loop}
A Higgs boson photoproduction in 
 gluon-gluon fusion via a quark loop,
\begin{equation}
gg \ra h(A),
\end{equation}
can be even more significant \cite{bk1,bk}.
The results for HERA with upgraded energy
are also presented in Fig. 1a.
A comparison of (2) with $\gamma g$ fusion (1)
  shows
that, for large $\tb$
and for mass  below $\sim 30$ GeV, the $gg$
fusion clearly dominates the total cross section.
Note that the total $gg$
cross section for $\tb$=30 is large: $\sigma \sim$  10$^5$ fb
for a Higgs mass of 5 GeV, falling to 5 fb at mass $\sim$
45 GeV (where  the $\sigma_{gg}$ meets the $\sigma_{\gamma g}$).

A mass of 45 GeV  corresponds to the lowest currently allowed
mass for MSSM Higgs bosons. (Note that in MSSM,
$h$ and $A$ tend to be degenerated in mass for large $\tb$).
Adding the contribution for $\tb$=30
from  processes (1) and (2) for both
scalar and pseudoscalar as well as that due to
WW fusion into $h$ \cite{hera96}
which are of the same order,
we estimate that HERA will produce 20-30 events of this
sort with luminosity 1 fb$^{-1}$.

In Fig.1a the $\tb$=1 case corresponds to
the  prediction for SM
Higgs production in $\gamma g$ and $gg$ fusion.
Applying the current limit for
the SM Higgs mass,  $M_{Higgs}\ge$
66 GeV \cite{hi}, we see how small
the corresponding $\gamma g$ and $gg$ cross sections are,
more than three  orders of magnitude
smaller than the rate found in
WW fusion into $h$ \cite{hera96}.

Comparing the {\it scalar} production
in Fig.1a with corresponding production in Fig.1b,
 in which the nominal HERA beam energy was used,
we see an extremely weak dependence on beam energy for {\it scalar} production.
In Fig.1b we also consider
pseudoscalar production
via gluon-gluon fusion (2)  for $\tb$=30.
The total cross sections  ($h(A)\ra all$)
and the cross sections for the $\tau^+ \tau^-$ final state are shown.
It is interesting to notice the large difference, in the
mass range below 10 GeV, between scalar and pseudoscalar
production seen both
in the total cross section $\sigma$
as well as in the   $\sigma \cdot $Br($\ra \tau^+ \tau^-)$.
Note that the difference almost disappears
 for the  $\tau^+ \tau^- $ cross section
above the $bb$ threshold.
\begin{figure}[ht] \label{fig:rap}
\vskip 4.3in\relax\noindent\hskip -1.15in
             \relax{\includegraphics{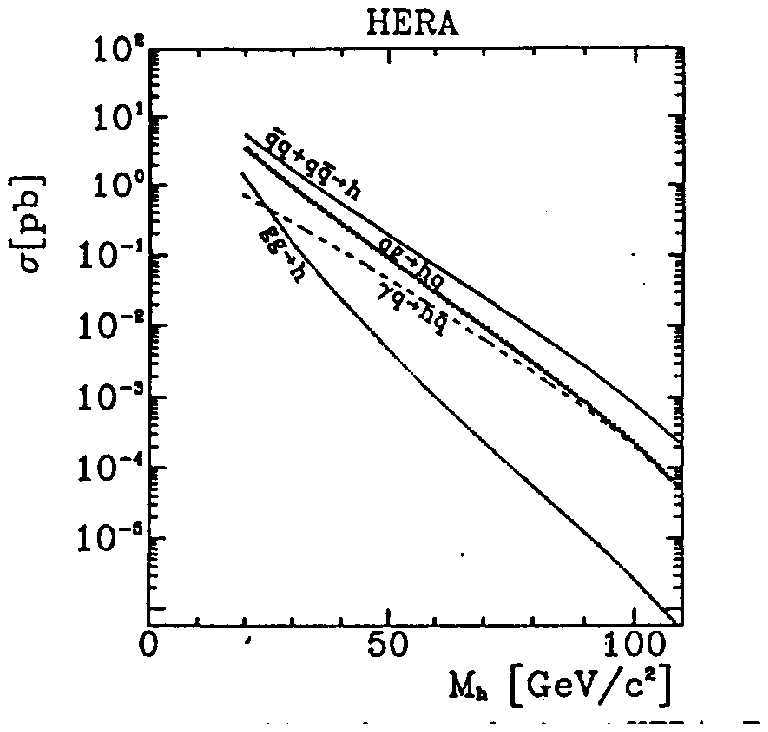}}
             \relax\noindent\hskip 3.45in
             \relax{\includegraphics{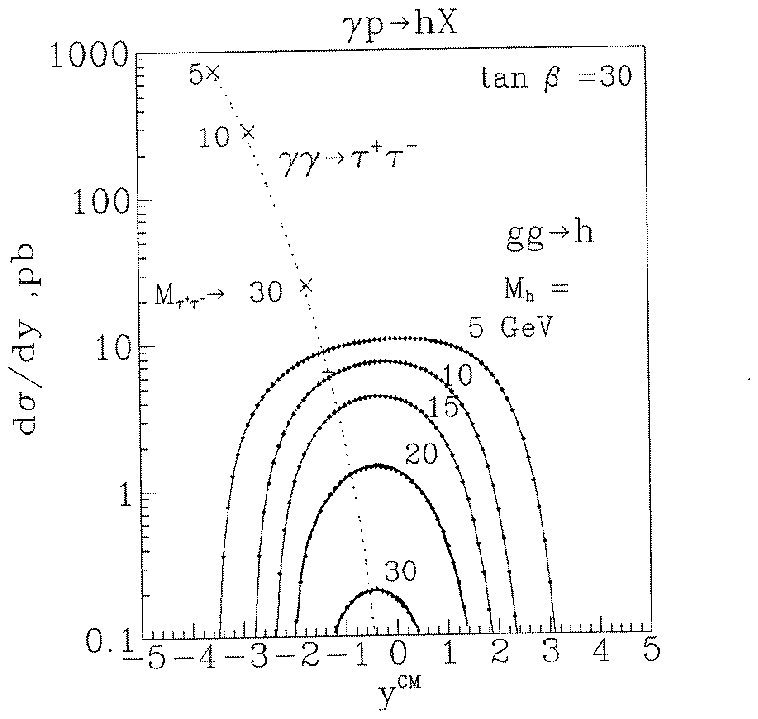}}
\vspace{-16.5ex}
\caption[junk]{ { \em
In the $ \gamma p$ CM system
at $ \sqrt{s_{ \gamma p} }$ =170 GeV a)
the cross section for scalar production in
various subprocesses  ($\tb$=20, from Ref.\cite{bk1})
and b) the rapidity distribution  for
 $ \tau^+ \tau^- $
pair are presented.
The background ($ \gamma \gamma \ra \tau^+ \tau^- $ )
and the signal (the scalar Higgs boson contribution,
integrated over
 $\Delta M_h $=1 GeV) are shown (from Ref.\cite{bk}).
  }}
\end{figure}

For detection,
it is useful to
study the rapidity distribution ${d \sigma }/{dy}$ of the Higgs bosons
in the $\gamma p$ centre of mass system.
Note that $y=-{{1}\over{2}}log{{E_h-p_h}\over{E_h+p_h}}
=-{{1}\over{2}}log{{x_{\gamma}}\over{x_p}}$, where $x_p(x_{\gamma})$
are the ratio of the energy of the gluon to the energy of the proton and photon,
respectively.
The (almost)
symmetric shape of the rapidity distribution found for the signal
is extremely useful
to reduce the
background and to separate the $gg\rightarrow h(A)$ contribution,
which we will discuss now for the $\tau^{+} \tau^{-}$ final state.

The main background in the mass range between
$\tau^{+} \tau^{-}$ and $b b$
thresholds is due to $\gamma\gamma\rightarrow \tau^+ \tau^- $.
In the region of negative rapidity
 ${d \sigma }/{dy_{\tau^+ \tau^-}}$ is very large,
\eg ~,for $\gamma p$ energy equal to 170 GeV
the cross section  $\sim$ 800 pb
at the edge of phase space ($y_{\tau^+ \tau^-}
\sim -4$), and it then falls rapidly when $y_{\tau^+ \tau^-}$
approaches $0$. At the same time, the signal reaches at most 10 pb
(for $M_h$=5 GeV). The results for a scalar  are shown in Fig.2b.
The region of positive rapidity  is {\it {not}} allowed
kinematically for this process since here one photon
interacts directly with
$x_{\gamma}=1$, and therefore $y_{\tau^+ \tau^-}
 =-{{1}\over{2}}log{{1}\over{x_p}}\leq 0$.
A significantly different topology found for
 $\gamma\gamma\rightarrow \tau^+\tau^-$ events
than for the signal should
allow a reduction of this background.

The other sources of background are
$q\bar q\rightarrow\tau^+\tau^-$ processes (not shown here).
These processes contribute to positive and negative
rapidity $y_{\tau^+ \tau^-}$, with a flat and
relatively low cross-section
(below 0.5 pb) in the central region.

Note that Higgs
decaying into $b$-quarks has a
much more severe background, and we will not discussed it here
(see Ref.\cite{bk1,bk}).

Assuming a luminosity ${\cal L}_{ep}$=250 pb$^{-1}$/y
we predict that $gg$ fusion
will produce around one thousand events
per annum for $M_h=5$ GeV (and roughly 10 events for
$M_h=30$ GeV).
A clear signature for the tagged case with a $\tau^+\tau^-$ final state
 at positive centre-of-mass
rapidities of the Higgs scalar should be seen, even for a Higgs mass
above the $bb$ threshold
(more details can be found in Ref.\cite{bk}).
For the pseudoscalar case even more events are expected
in the mass region below 10 GeV.

To show the potential of HERA, the exclusion plot
based on $gg$ fusion via a quark loop with the
$\tau^+ \tau^-$ final state
can be drawn. In this case, as we mentioned above,
it is easy to find the
part of phase space where the background is hopefully negligible.
To calculate the 95\% C.L. for allowed value of $\tb$
we take into account signal events corresponding only to
the positive rapidity region (in the $\gamma p$ CM system).
 The  results
for the $ep$ luminosity ${\cal L}_{ep}$
=25 pb$^{-1}$ and 500 pb$^{-1}$
are presented in Fig. 3 and will be discused in the following section.

\section{Exclusion plot for 2HDM}
In Fig.3 the  95\% C.L. exclusion curves for the $\tb$
in the general  2HDM ("Model II")
obtained by us  for a light scalar (solid lines)
and for a light pseudoscalar (dashed lines)
are presented in mass range below 40 GeV.
For comparison results from LEP I analysis presented recently
by ALEPH collaboration 
for pseudoscalar \cite{alef} is also shown (dotted line).
The region of ($\tb, M_{h(A)}$) above curves is excluded.
\begin{figure}[ht] \label{fig:excl}
\vskip 4.in\relax\noindent\hskip -0.50in
             \relax{\includegraphics{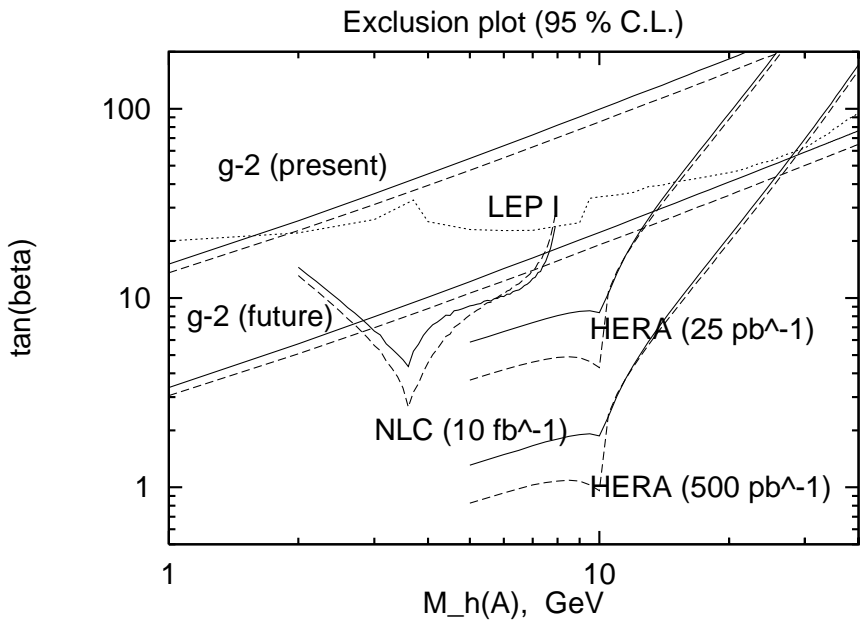}}
\vspace{-12.5ex}
\caption[junk]{ { \em
The 95\% C.L. exclusion plot for a light
scalar(solid lines) or light pseudoscalar (dashed lines)
in 2HDM.
The limits {derivable} from present
$(g - 2)_{\mu}$ measurements and from existing
LEP I results (pseudoscalar production in the
Yukawa process)
(dotted line) are shown. The possible exclusions from HERA
(the gluon-gluon fusion via a quark loop with the
$\tau^+\tau^-$ final state)
for luminosity 25 pb$^{-1}$ and 500 pb$^{-1}$ as well from
$\gamma \gamma \rightarrow \mu^+ \mu^-$ at low energy NLC
(10 fb$^{-1}$) are also presented (from \cite{deb12}).
Possible limits from future
data for $(g-2)_{ \mu}$ are also shown.
The parameter space above the curves can be ruled out.
  }}
\end{figure}

Constraints on $ \tb$
were obtained from the
existing  $(g-2)_{\mu}$ data, including LEP I mass limits
(Sec.2.2).
We see  that already the present   $(g-2)_{\mu}$ data
 improve LEP I limits
 on $\tb$ for $M_A \lsim $ 2 GeV.
A similar situation  should  hold for a 2HDM with a light scalar,
although here the Yukawa process may be more restrictive for $M_h\le$
10 GeV \cite{kk}.

The
future improvement
in the accuracy  by a factor of 20 in the  forthcoming
$(g-2)_{\mu}$ experiment  may lead to more stringent limits
than provided by LEP I up to a mass of $h$ or $A$
equal to 30 GeV,
if the mass difference between scalar and pseudoscalar is $\sim M_Z$,
or to even higher mass for a larger mass difference \cite{g22}.
Note, however, that there is some arbitrariness in the deriving the
expected bounds for the $\delta a_{\mu}^{new}$.

The search at HERA in the gluon-gluon fusion via a quark loop
may lead to even more stringent
limits (see Fig.3) for the mass range 5--15 (5--25) GeV, provided
the luminosity will reach 25 (500) pb$^{-1}$ and the
efficiency for the $\tau^+ \tau^-$ final state will be high
enough \footnote{In this analysis the 100\% efficiency
has been assumed. If the efficiency will be 10 \% the corresponding
limits will be larger by factor 3.3 (a simple scaling) for fixed luminosity.
}.
The other production mechanisms, such as $\gamma g$ fusion
and other subprocesses with the resolved photon, are expected to improve
these limits further \cite{bk1,bk,deb9}.

In the very low mass range,
additional limits can be obtained from the low energy
NL $\gamma \gamma$ collider with $\sqrt {s_{ee}}$=10 GeV.
In Ref.\cite{deb12}
we found that the exclusion based on $\gamma \gamma$ fusion
into Higgs, decaying into $\mu^+ \mu^-$,
 may be very efficient.
In Fig.3 the
results corresponding to
a luminosity  10 fb$^{-1}$ are presented.

\section {Conclusion}
In the framework of 2HDM,
a light neutral Higgs scalar or pseudoscalar
in the mass range below 40-50 GeV
is not ruled out by the present LEP I and $(g-2)_{\mu}$ data.
The other low energy experiments
cover only part of parameter space of 2HDM; some, such as the Wilczek process,
have large
theoretical uncertainties
both due to the QCD and relativistic corrections(\cite{wil,hunter})(see also
discussion
in \cite{bk,ames}).

The role of the forthcoming $(g-2)_{\mu}$
measurement seems to be crucial in
clarifying which scenario of 2HDM is allowed:  with light scalar or
with a light pseudoscalar.
If the $\delta a_{\mu}^{new}$ is positive(negative) then the light
pseudoscalar(scalar) is excluded.
Then, further constraints on the coupling of the allowed light Higgs
particle can be obtained from
HERA, which is very well suited for this.
The simple estimation
performed at luminosity 500 pb $^{-1}$ for
one particular production mechanism,
namely gluon-gluon fusion, is already promising;
adding more processes may further improve the situation significantly.
The most important experimental handle is a good efficiency for the
$\tau^+\tau^-$ channel.

All this suggests the large  discovery/exclusion
potential of HERA
for the mass range 5--20 GeV \cite{bk}.
It is unlikely that the LEP/LHC experiments
will have a larger potential in such a mass region \cite{bk1}.

The very low mass region may also be studied
at low energy 
NLC machines.
We found that the exclusion based on $\gamma \gamma$ fusion
into Higgs, decaying into $\mu^+ \mu^-$,
may be very efficient
in probing the Higgs sector of 2HDM, even for luminosity 100 pb$^{-1}$.
It is not clear however if these
low energy options will come into operation.

Future experiments will clarify the status of the
general 2HDM with a light neutral Higgs particle --
the role of HERA in such a study may be very important.

By contrast, for the MSSM
the potential of HERA even with  luminosity
1 fb$^{-1}$ is relatively poor, producing only 20-30 events
of $h$ and $A$, if the mass is in the 45-50 GeV range
for $\tb$=30.
\section{Acknowledgements}
I would like to thank D. Choudhury and J.\.Zochowski
for important contribution to
this paper and  B. Kniehl for  fruitful discussions.
I am grateful very much to P. Zerwas for very important
suggestion and support. I am also grateful  to
H. Dreiner, S. Ritz and U. Katz for very importrant
comments and suggestions.
Supported in part by the Polish Committee for Scientific Research.

\end{document}